\documentclass[aps,pra,superscriptaddress,preprintnumbers,twocolumn,showpacs]{revtex4}
\usepackage{amsmath}
\usepackage{amssymb}
\usepackage{graphicx}
\usepackage{color}
\usepackage{bm}

\renewcommand{\Re}{\mathop{\text{Re}}\nolimits}

\newcommand{\Tr}{\mathop{\text{Tr}}\nolimits}
\newcommand{\ket}[1]{|{#1}\rangle}
\newcommand{\bra}[1]{\langle{#1}|}

\definecolor{dgreen}{rgb}{0,0.5,0}

\definecolor{delete}{cmyk}{0.5,0,0,0}
\definecolor{deletey}{cmyk}{0,0.5,0,0}

\date{\today}
\begin{document}
\title{Distillation by repeated measurements: continuous spectrum case}
\author{Bruno Bellomo}
\author{Giuseppe Compagno}
\affiliation{CNISM \& Dipartimento di Scienze Fisiche ed Astronomiche,
Universit\`{a} di Palermo, via Archirafi 36, 90123 Palermo, Italy}
\author{Hiromichi Nakazato}
\affiliation{Department of  Physics, Waseda University, Tokyo 169-8555, Japan}
\author{Kazuya Yuasa}
\affiliation{Waseda Institute for Advanced Study, Waseda University, Tokyo 169-8050, Japan}

\begin{abstract}
Repeated measurements on a part of a bipartite system strongly affect the other part not measured, whose dynamics is regulated by an effective contracted evolution operator. When the spectrum of this operator is discrete, the latter system is driven into a pure state irrespective of the initial state, provided the spectrum satisfies certain conditions.
We here show that even in the case of continuous spectrum an effective distillation can occur under rather general conditions. We confirm it by applying our formalism to a simple model.
\end{abstract}

\maketitle

Distillation procedures aiming at driving quantum systems into pure states are relevant tools in the field of quantum information and computation \cite{ref:QIT}.
In fact, they can be exploited to control the state of a quantum system and play a crucial role to initialize quantum systems. Various purification schemes have been proposed \cite{ref:Purification}
and among them is a state generation strategy based on the extraction of a state through repeated measurements  \cite{Nakazato2003}.

Indeed, for a generic bipartite quantum system consisting of two interacting parts $X$ and $A$, repeated measurements on one part ($X$) can strongly affect the dynamics of the other ($A$).
In the case of measurements projecting $X$ in its initial state, the dynamics of $A$ is governed by an effective evolution operator $\hat{V}_{\tau}$ (introduced below), and it has been shown that, when the spectrum of this operator is such that its largest (in magnitude) eigenvalue is unique, discrete, and nondegenerate, then the non-measured system $A$ is driven toward a pure state irrespective of its initial condition \cite{Nakazato2003}.
This distillation procedure has been also utilized to produce entangled states of a multipartite system $A$ \cite{ref:qpf-ent},
in particular allowing to establish entanglement between two spatially separated systems via repeated measurements on an entanglement mediator \cite{ref:qpf-separate}.

The above requirements for distillation cannot be satisfied if the spectrum of $\hat{V}_{\tau}$ is continuous.
Such a situation can be found when one or both of the subsystems have a continuous spectrum.
It has been shown that, although the measured part $X$ has a continuous spectrum, $\hat{V}_{\tau}$ may have a discrete spectrum, provided the spectrum of the non-measured part $A$ is discrete \cite{ref:qpf-bruno}.
In this paper, we investigate the case where $\hat{V}_{\tau}$ itself is characterized by a continuous spectrum and show that even in such a case an effective distillation can happen under rather general conditions.

We prepare $X$ at time $t=0$  in a pure state $|\Phi\rangle$, e.g.\ by projecting $X$ to this state by a measurement, while $A$ is in an arbitrary mixed state $\hat{\rho}_A(0)$.
The unitary dynamics of the total system $X$+$A$, governed by the time-evolution operator $\hat{U}(\tau)$, is interrupted by the measurements performed
on $X$ at intervals $\tau$.
The measurements are so designed to project $X$ onto $\ket{\Phi}$.
The action of each measurement is represented by the projection operator
\begin{equation}
\hat{\mathcal{O}} =
|\Phi\rangle \langle \Phi |\otimes \mathbf{1}_A.
\end{equation}
The state of the total system after $N$ measurements is thus described by
\begin{equation}\label{total density matrix evolution}
     \hat{\rho}_T(N)\propto [\hat{\mathcal{O}}\hat{U}(\tau)]^N[|\Phi\rangle \langle \Phi|\otimes   \hat{\rho}_A(0)]
      [ \hat{U}^{\dagger}(\tau)\hat{\mathcal{O}}]^N .
\end{equation}
Following \cite{Nakazato2003}, we introduce the projected evolution operator between two consecutive measurements,
\begin{equation}\label{contracted evolution operator}
    \hat{V}_\tau= \langle \Phi  | \hat{U}(\tau) |\Phi\rangle,
\end{equation}
so that, after \textit{N} measurements on $X$,
system $A$ is described by the density matrix
\begin{gather}
     \hat{\rho}_A(N)=\hat{V}_\tau^N\hat{\rho}_A(0)
      \hat{V}_\tau^{\dagger N}/P(N),
\label{field density matrix evolution}
\\
    P(N)=\Tr_{A}\{\hat{V}_{\tau}^N\hat{\rho}_A(0)
      \hat{V}_\tau^{\dagger N} \},
\label{probability}
\end{gather}
where the normalization factor $P(N)$ represents the survival probability that $X$ is always found (up to $N$ measurements) in the state $\ket{\Phi}$ by every measurement and thus gives the probability to obtain the state (\ref{field density matrix evolution}).

Since the operator $\hat{V}_{\tau}$ is not Hermitian, $ \hat{V}_{\tau} \neq  \hat{V}_{\tau}^\dag$, in general, we need to set up both the right- and left-eigenvalue problems, $\hat{V}_{\tau}\ket{u_E}=\lambda_E\ket{u_E}$ and $\bra{v_E}\hat{V}_{\tau}=\lambda_E\bra{v_E}$.
Let us assume that the spectrum of the operator $\hat{V}_{\tau}$ is continuous and nondegenerate, and its eigenvectors form a complete orthonormal set in the following sense: $\int dE\, \ket{u_E}\bra{v_E}=\mathbf{1}_A$ and $\langle v_E\ket{u_{E'}}=\delta(E-E')$.
The operator $\hat{V}_{\tau}$ is expanded in terms
of these eigenvectors as
\begin{equation}\label{diagonal v tau}
     \hat{V}_{\tau}=\int d E\, \lambda_E \ket{u_E}\bra{v_E},
\end{equation}
and $A$ is driven by the operator
\begin{equation}
\hat{V}_{\tau}^N=\int d E\, \lambda_E^N  \ket{u_E}\bra{v_E}
\end{equation}
as the measurement is repeated.

In order to discuss the possibility to obtain an effective distillation of $A$, we will compare the survival probability $P(N)$ and the purity of $A$, quantified by
\begin{equation}
\Pi(N)=\Tr_{A}\{\hat{\rho}_A^2(N)\},
\end{equation}
as functions of the number of measurements $N$. We endeavor to find conditions under which an effective distillation of $A$, in the sense clarified later, can be achieved and consider the cases where the right-eigenvectors are orthogonal to each other $\langle u_E \ket{u_{E'}}=\delta(E-E')$ (which is actually the case in the examples studied below). In such cases, the survival probability (\ref{probability}) takes the form
\begin{equation}\label{probability computed}
    P(N)=\int  d E\,  |\lambda_E|^{2N}  \bra{u_E} \hat{\rho}_A(0) \ket{u_E},
\end{equation}
and the purity of $A$ is expressed as
\begin{align}
\Pi(N)
 =\frac{1}{P^2(N)}\int dE\, d E'\,  &|\lambda_E|^{2N}|\lambda_{E'}|^{2N}
 \nonumber\\[-1mm]
 &{}\times|\bra{u_E}\hat{\rho}_A(0)\ket{u_{E'}}|^2.
\label{purity}
\end{align}
Let $E_{*}$ be the value of $E$ at which $|\lambda_E|$ has its (unique) absolute maximum, $|\lambda_{E}|'_{E=E_*}=0$, and consider the Taylor expansion of $\Lambda(E)=-\ln |\lambda_E|^2$ around $E_{*}$, $|\lambda_E|^{2N}=e^{N \ln |\lambda_E|^2}\approx e^{-N \Lambda(E_{*})-N\Lambda''(E_{*})(E-E_{*})^2/2+\cdots}$.
Notice that the higher order terms become less relevant as $N$ increases, and $|\lambda_E|^{2N}$ becomes well approximated by a Gaussian
\begin{gather}
    |\lambda_E|^{2N}\approx f(N)\frac{1}{\sqrt{2 \pi\Delta_N^2}}e^{-(E-E_{*})^2/2\Delta^2_N},
    \nonumber\\
    f(N)=  \sqrt{2 \pi\Delta_N^2}e^{-N \Lambda(E_{*})}, \quad   \Delta_N=\frac{1}{\sqrt{N\Lambda''(E_{*})}}.
\label{Gaussian approximation}
\end{gather}
Observe here that the Gaussian in (\ref{Gaussian approximation}) becomes narrower like $\Delta_N\propto1/\sqrt{N}$ as $N$ increases, and a narrow band around $E_*$ is filtered in the spectrum.
Putting $x=(E-E_{*})/\Delta_N$, the survival probability is shown to behave for large $N$ as
\begin{align}
    P(N)&\approx f(N)\int\frac{dx}{\sqrt{2\pi}}\,e^{-x^2/2}\bra{u_{E_{*}+x\Delta_N}}\hat{\rho}_A(0)\ket{u_{E_{*}+x\Delta_N}}\nonumber\\
&\approx f(N)\left(
g(0)
+\frac{1}{2}g''(0)\Delta^2_N\right),
   \label{limit survival probabilit2}
\end{align}
where $g(y)\equiv{}\bra{u_{E_{*}+y}}\hat{\rho}_A(0)\ket{u_{E_{*}+y}}$, and the purity given by (\ref{purity}) becomes
\begin{align}
\Pi(N)
\approx {}& \frac{f^2(N)}{P^2(N)}\int\frac{d x\, d x'}{2\pi}\, e^{-(x^2+{x'}^2)/2}
\nonumber \\[-1truemm]
&\qquad\qquad\qquad{} \times
 |\bra{u_{E_{*}+x\Delta_N}}\hat{\rho}_A(0)\ket{u_{E_{*}+x'\Delta_N}}|^2
 \nonumber\\[1truemm]
\approx{} &1-\Delta_N^2\frac{g(0)g''(0)-h_{yy}(0,0)}{g^2(0)}.
\label{final limit purity}
\end{align}
Here we have introduced $h(y,y')=|\bra{u_{E_{*}+y}}\hat{\rho}_A(0)\times\ket{u_{E_{*}+y'}}|^2$ and denoted its second partial derivative with respect to $y$ as $h_{yy}(y,y')$. One sees that, for large $N$, the purity $\Pi(N)$ approaches 1 with a rate determined by $\Delta^2_N \propto1/N$. System $A$ is thus asymptotically purified toward $\ket{u_{E_{*}}}$ by the repeated measurements on $X$.

In this process, however, one should pay attention to the behavior of the survival probability: the purity $\Pi(N)$ should reach $1$ quick enough before the probability $P(N)$ decays out completely.
The comparison between the rates of the decay of $P(N)$ to 0 and the approach of $\Pi(N)$ to 1 leads to the following optimization criterion in order to obtain an efficient distillation of the pure state $\ket{u_{E_{*}}}$.
If $\Lambda(E_{*})= 0$, i.e.\ $|\lambda_{E_{*}}|^{2}=1$, the exponential decay $e^{-N\Lambda(E_{*})}$ in $ P(N)$ disappears and the decay of $P(N)$ is ruled by $\Delta_N \propto 1/\sqrt{N}$. This decay is slower than the approach of the purity $\Pi(N)$ to $1$, i.e.\ $\Delta^2_N \propto 1/N$. Therefore, if the magnitude of the eigenvalue associated to the eigenstate $\ket{u_{E_{*}}}$ is $|\lambda_{E_{*}}|=1$, system $A$ is driven toward the pure state $\ket{u_{E_{*}}}$ with a rate faster than the decay of the survival probability $P(N)$.

Remark that, when $\langle u_E \ket{u_{E'}}=\delta(E-E')$ is not satisfied, it can be shown that $P(N)$ and $\Pi(N)$ at best decrease and increase respectively as $\Delta_N^2\propto1/N$ for large $N$, so that distillation seems more difficult to achieve.

\textbf{Model.}
We now apply, as an example, the above framework to a specific model.
We consider a particle of mass $m$ interacting with a single cavity mode of frequency $\omega$.
This system has been studied in \cite{ref:qpf-bruno}
to analyze how the repeated measurements on the particle affect the dynamics of the cavity mode.
In that case, although the measured part (particle) has a continuous spectrum, the effective dynamics of the cavity mode is described by an operator $\hat{V}_{\tau}$ characterized by a discrete spectrum.
Here, the opposite case is investigated, that is, the cavity mode is repeatedly projected onto its initial state by measurements. In this case, as we will show, the effective dynamics of the particle is described by an operator $\hat{V}_{\tau}$ having a continuous spectrum.

The Hamiltonian describing the system is
\begin{equation}\label{hamiltoniana di partenza}
  \hat{H}=\frac{\hat{p}^2}{2m}+\hbar\omega
 \left( \hat{a}^{\dag}\hat{a}+\frac{1}{2}\right)+
   g \hat{p}(
  \hat{a}^{\dag}
  +\hat{a} ),
\end{equation}
where $\hat{p}$ is the momentum operator of the particle,
$\hat{a}$ and $\hat{a}^{\dag}$ are the
annihilation and creation operators of the cavity mode, respectively, satisfying
the commutation rule
$[\hat{a},\hat{a}^{\dag}]=1$, and the real parameter $g$ is the
coupling constant.
The dynamics described by the Hamiltonian (\ref{hamiltoniana di partenza}) is exactly solvable, and the exact evolution operator at time $\tau$ in the Schr\"{o}dinger picture is given by \cite{ref:ExactSolution,ref:qpf-bruno}
\begin{align}\label{evolution operator}
   \hat{U}(\tau)={} &e^{-i\xi_\tau\hat{p}^2 \tau/2 m\hbar}
e^{-i \omega \tau( \hat{a}^{\dag}\hat{a}+1/2)}
e^{\hat{p} (g_\tau\hat{a}^{\dag} -g_\tau^*\hat{a})},
\end{align}
where
$\xi_\tau =1 -(2 mg^2/\hbar\omega)[1-(\sin \omega\tau)/ \omega\tau]$ and $ g_\tau =g(1-e^{i\omega \tau})/\hbar \omega$.

The projected evolution operator $\hat{V}_\tau$ defined by (\ref{contracted evolution operator})
strongly depends on the choice of the state $\ket{\Phi}$ on which the cavity mode is repeatedly projected at intervals $\tau$.
Nevertheless, it results to be diagonal in the momentum representation for any choice of $\ket{\Phi}$.
It follows that, for any choice of the cavity state $\ket{\Phi}$, the operator $\hat{V}_{\tau}^N$ has a continuous spectrum of the form $\hat{V}^N_\tau =\int dp\, \lambda_p^N\ket{p}\bra{p}$,
where $\lambda_p$ is the eigenvalue whose explicit form depends on the choice of the cavity state $\ket{\Phi}$ to be measured.
In the following, we examine how the above general framework for distillation with a continuous spectrum of $\hat{V}_\tau$ works, by looking at two cases with different cavity states $\ket{\Phi}$.
In the first case, $\ket{\Phi}$ is assumed to be a coherent state $\ket{\alpha}$, while in the second, a number state $\ket{n}$.
Both cases are of interest from an experimental point of view \cite{ref:CQED}.
The first choice leads directly to a Gaussian form of $|\lambda_p|^{2N}$, while in the second case, $|\lambda_p|^{2N}$ itself is not Gaussian, but it becomes well approximated by a Gaussian, and we will see that the above general framework actually works.

As the initial state of the particle, we take a generic Gaussian state  $\hat{\rho}_{A}(0)=\int dp\,dp'\,\ket{p}\rho_{pp'}(0)\bra{p'}$ with
\begin{align}
\label{generic gaussian state}
     \rho_{pp'}(0)=\frac{1}{\sqrt{2\pi(\Delta p)_0^2}}
     &e^{-\frac{(p+p'-2p_0)^2}{8(\Delta p)_0^2}
          -\frac{(p-p')^2}{8(\Delta p)_0^2\Pi_0^2}-\frac{i}{\hbar}(p-p')x_0
      }
      \nonumber\\[-2mm]
      &{}\times e^{-\frac{iB}{4(\Delta p)_0^2\Pi_0}(p-p')(p+p'-2p_0)},
\end{align}
where $B=\sqrt{4(\Delta x)_0^2(\Delta p)_0^2\Pi_0^2/\hbar^2-1}$, $p_0$ and $x_0$, and $(\Delta p)_0^2$ and $(\Delta x)_0^2$ are the averages and variances of the momentum and the position, and $\Pi_0$ the purity of this initial state \cite{Joos2002}.

\textbf{Cavity coherent state
 $\ket{\Phi}=|\alpha\rangle $ ($\alpha=|\alpha|e^{i \gamma}$).}
 In this case, the projected evolution operator reads
 \begin{align}
\label{V diagonal}
    \hat{V}_\tau
    ={}& e^{-i\omega  \tau/2 -2 i |\alpha|^2\sin(\omega \tau/2)e^{-i\omega\tau/2}}
    \nonumber \\[-1mm] &{} \times
 \int dp\,\ket{p}\bra{p}
 e^{-i\xi_\tau p^2 \tau/2 m\hbar -p^2|g_\tau|^2/2+pb} ,
\end{align}
where $ b=-\alpha g_\tau^*+e^{-i \omega \tau} g_\tau \alpha^*$.
The survival probability defined in (\ref{probability}) takes the form
\begin{align}\label{survival probability computed}
    \!P(N)={}&\frac{e^{-4N|\alpha|^2\sin^2\!\gamma \sin^2(\omega \tau/2)}}{{\sqrt{1+2 |g_\tau|^2  (\Delta p)_{0}^2 N}}}
    \nonumber \\
    &{}\times\exp\!\left(
    -\frac{(p_0-\Re b/|g_\tau|^2)^2}{2(\Delta p)_{0}^2\left[1+1/2 |g_\tau|^2  (\Delta p)_{0}^2 N\right]}\right),
\end{align}
where $\Re b= -4(g|\alpha | /\hbar \omega)   \sin^2(\omega \tau/2)\cos\gamma$.
In the limit of large number of measurements, $ N \gg 1 / 2 |g_\tau|^2 (\Delta p)_{0}^2 $, $P(N)$ reduces to
\begin{align}
P(N)\approx{}& \frac{e^{-4N|\alpha|^2\sin^2\!\gamma \sin^2(\omega \tau/2)}}{{\sqrt{2 |g_\tau|^2  (\Delta p)_{0}^2 N}}}
    \nonumber \\
    &{}\times\exp\!\left[
    -\frac{1}{2(\Delta p)_{0}^2}\left(p_0-\frac{\Re b}{|g_\tau|^2}\right)^2
    \right].
\end{align}
This formula corresponds to (\ref{limit survival probabilit2}), without the second-order term proportional to $\Delta_N^2$, through the relationships
\begin{align}
&\Delta_N = \frac{1}{\sqrt{2|g_\tau|^2N}}, \qquad p_{*}= \frac{\Re b}{|g_\tau|^2},
\nonumber \\
&f(N)=\sqrt{2\pi\Delta_N^2}
\,e^{-4N|\alpha|^2\sin^2\!\gamma \sin^2(\omega \tau/2)},
  \nonumber\displaybreak[0] \\
&\bra{p_{*}}\hat{\rho}_A(0)\ket{p_{*}}
=\frac{1}{\sqrt{2 \pi(\Delta p)_{0}^2}}
e^{-(p_{*}-p_0)^2/2(\Delta p)_{0}^2}.
\end{align}
One can see that the selected momentum $p_{*}$, being proportional to $\Re b$ given below (\ref{survival probability computed}), can be chosen by tuning the parameters properly.
In particular for $\gamma=0,\pi$, the exponential decay in the survival probability, i.e.\ in $f(N)$, can be suppressed, being the final momentum $p_{*}$ negative in the first case $\gamma=0$ and positive in the second $\gamma=\pi$.
The other parameters can be tuned in order to obtain the desired modulus of the final momentum $p_*$.
Of course, the farther it is from the initial average momentum $p_0$, the smaller the exponential factor in $\bra{p_*}\hat{\rho}_A(0)\ket{p_*}$ will be, since it depends on $p_*-p_0$.

Let us also look at the evolution of the purity:
\begin{equation}
\label{coherent purity evolution}
    \Pi(N)=\sqrt{\frac{1+2|g_\tau|^2(\Delta p)_0^2 N}{1/\Pi_0^2+2|g_\tau|^2(\Delta p)_0^2 N}}
    \approx 1-\frac{1/\Pi_0^2-1}{2(\Delta p)_0^2}\Delta_N^2.
\end{equation}
If the initial state of the particle is pure, $\Pi_0=1$, it remains pure after each measurement, while in general, the purity of the particle increases as the measurements go on. This formula is again consistent with (\ref{final limit purity}).

\begin{figure}[b]
\begin{center}
\includegraphics[width=0.76\columnwidth]{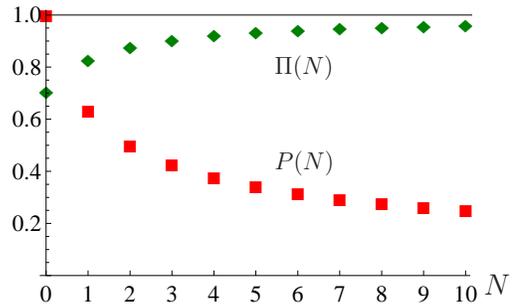}
\caption{(Color online) $P(N)$ (squares) vs.\ $\Pi(N)$ (diamonds) as functions of the number of measurements $N$, when the coherent state $\ket{\alpha}$ of the cavity mode is repeatedly measured. The parameters are $\omega\tau=\pi/4$, $\sqrt{m/\hbar\omega}\,g=5$,  $p_0/\sqrt{m\hbar\omega}=0.1$, $(\Delta p)_0/\sqrt{m\hbar\omega}=0.2$, $\Pi_0=1/\sqrt{2}$, and $|\alpha|=1$, $\gamma=\pi$. This plot is independent of $x_0$ and $(\Delta x)_0$. The final selected momentum is $p_*=2p_0$.}
\label{fig:survivals}
\end{center}
\end{figure}

In Fig.\ \ref{fig:survivals}, the probability $P(N)$ in (\ref{survival probability computed}) and the purity $\Pi(N)$ in (\ref{coherent purity evolution}) are plotted for $\gamma=\pi$.
The purity approaches 1 when the probability is yet far from 0.

\begin{figure}[t]
\begin{center}
\includegraphics[width=0.76\columnwidth]{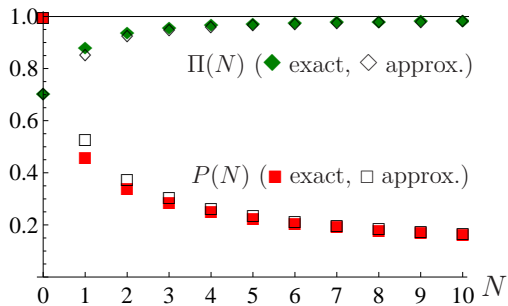}
\caption{(Color online) Exact (filled marks) and approximate (empty marks) behaviors of $P(N)$ (squares) vs.\ $\Pi(N)$ (diamonds) as functions of the number of measurements $N$, when a number state $\ket{1}$ of the cavity mode is repeatedly measured. The parameters are the same as in Fig.\ \ref{fig:survivals}, while $p_*=0$.}
\label{fig:number survivals}
\end{center}
\end{figure}

\textbf{Cavity number state $\ket{\Phi}=|n \rangle$.}
For $n=1$, $\hat{V}_\tau$ results in
\begin{align}
\label{nuno}
\hat{V}_\tau
    =
    \int dp\,\ket{p}\bra{p}&
 (1-p^2|g_\tau|^2)e^{-3i\omega\tau/2}
    \nonumber\\[-1mm]
    &
    {}\times
 e^{-i\xi_\tau p^2 \tau/2 m\hbar-p^2|g_\tau|^2/2}.
\end{align}
In this case, our generic formulas in (\ref{Gaussian approximation}) suggest
\begin{gather}
\Delta_N = \frac{1}{\sqrt{6|g_\tau|^2N}} , \quad p_{*}= 0, \quad f(N) =\sqrt{ \frac{\pi}{3|g_\tau|^2N}},
  \nonumber\displaybreak[0] \\
\bra{p_{*}}\hat{\rho}_A(0)\ket{p_{*}}=\frac{1}{\sqrt{2 \pi(\Delta p)_{0}^2}}
e^{-p_0^2/2(\Delta p)_{0}^2},
\end{gather}
and the asymptotic formula (\ref{limit survival probabilit2}) for the survival probability $P(N)$ for a large number of measurements yields
\begin{equation}\label{number survival probability}
  P(N)\approx \frac{1}{\sqrt{6|g_\tau|^2(\Delta p)_{0}^2 N} }
e^{-p^2/2(\Delta p)_{0}^2}.
\end{equation}
On the other hand, (\ref{final limit purity}) gives the asymptotic behavior of the purity $\Pi(N)$ for a large number of measurements,
\begin{equation}
\label{number purity evolution}
    \Pi(N)\approx 1-\frac{1/\Pi_0^2-1}{2(\Delta p)_0^2}\Delta_N^2,
\end{equation}
which is the same as that obtained in the coherent state case in (\ref{coherent purity evolution}).
These quantities, the survival probability $P(N)$ in (\ref{number survival probability}) and the purity $\Pi(N)$ in (\ref{number purity evolution}), are plotted in Fig.\ \ref{fig:number survivals}, compared with the exact results computed numerically on the basis of the expression for $\hat{V}_\tau$ in (\ref{nuno}).
It shows that the exact results and the asymptotic formulas match already after a small number of measurements and that the purity $\Pi(N)$ approaches 1 before the survival probability $P(N)$ decays to 0.

\textbf{Conclusions.}
We have studied the distillation process, in which one part of a bipartite system is purified by repeatedly projecting the other part onto a certain state,
in the case where the dynamics is regulated by an operator $\hat{V}_\tau$ characterized by a continuous spectrum. When the maximum of the continuous spectrum $|\lambda_E|$ is unique at $E_*$ and the second derivative $\Lambda''(E_*)\neq 0$ exists, the spectrum $|\lambda_E|^{2N}$ becomes Gaussian around $E_*$ as the number of measurements $N$ increases, and the state is purified to $\ket{E_*}$. This purification is optimized, if it is allowed to tune the parameters so that $|\lambda_E|=1$, by which the purity increases faster than the decrease of the survival probability.
The distillation by repeated measurements has been considered to be possible only if $\hat{V}_\tau$ has a discrete spectrum.
The present analysis reveals that this procedure can be applied to a wider class of systems.

B.B.\ thanks people at Waseda University for their hospitality.
This work is supported by a Special Coordination Fund for Promoting Science and Technology, and a Grant-in-Aid for Young Scientists (B), both from MEXT, Japan,
by a Grant-in-Aid for Scientific Research (C) from JSPS, Japan,
by a bilateral Italian-Japanese Project of MIUR, Italy, and
by a Joint Italian-Japanese Laboratory of MAE, Italy.

\end{document}